\documentstyle[12pt]{article}

\newcommand{\be}{\begin{equation}}
\newcommand{\ee}{\end{equation}}
\newcommand{\Dlt}{\Delta}
\newcommand{\dlt}{\delta}
\newcommand{\prt}{\partial}
\newcommand{\br}{{\bf r}}
\newcommand{\bk}{{\bf k}}
\newcommand{\bt}{\beta}

\newcommand{\ep}{\varepsilon}

\newcommand{\ra}{\rightarrow}
\newcommand{\sgm}{\sigma}

\newcommand{\om}{\omega}
\newcommand{\Om}{\Omega}

\newcommand{\dgr}{\dagger}

\begin{document}

\begin{center}

{\Large {\bf Structure factor of Bose-condensed systems} \\ [5mm]

V.I. Yukalov} \\ [3mm]

{\it Bogolubov Laboratory of Theoretical Physics, \\
Joint Institute for Nuclear Research, Dubna 141980, Russia}

\end{center}

\begin{abstract}

The structure factor for a Bose system with Bose-Einstein condensate is
considered in the frame of the self-consistent mean-field approximation.
The accomplished analysis demonstrates the principal importance of the
following three points: the necessity of preserving the approximation
order, the necessity of taking into account anomalous averages, and the
necessity of gauge symmetry breaking. If any one of these necessary
conditions is not satisfied, calculations yield the appearance of
unphysical divergences of the structure factor, which implies the
occurrence of fictitious instability, which contradicts experiments.

\end{abstract}

\vskip 1cm


\vskip 1cm


\newpage

\section{Introduction}

The structure factor is an important quantity characterizing, as follows
from its name, the structural properties of the considered system as well
as the spectrum of collective excitations. The structure factor can be
measured in scattering experiments. It is also related to other directly
measurable quantities, such as the isothermal compressibility and sound
velocity.

The structure factor of Bose systems has been repeatedly investigated
for the case of liquid $^4$He, both theoretically [1,2] and experimentally
[3--9]. In recent years, the physics of dilute Bose gases [10] has attracted
a great deal of attention (see review articles [11--15]). The structure
factor of cold trapped atoms was also measured [16]. Contrary to liquid
helium, for which strong interparticle correlations must be accurately taken
into account [2,17], dilute gases allow for their description the use of the
mean-field approach. Thus, the measured structure factor of dilute trapped
$^{87}$Rb atoms was found [16] to be in good agreement with the Bogolubov
approximation [18,19]. The latter is known to be well justified for weakly
nonideal Bose gases, with asymptotically small interactions. When atomic
interactions are not asymptotically weak, but rather are finite, their
theory encounters the notorious Hohenberg-Martin [20] dilemma of conserving
versus gapless approaches. This dilemma has recently been resolved by
constructing a representative statistical ensemble for Bose systems with
broken gauge symmetry [21,22] and by developing a self-consistent mean-field
approximation [23--26].

The aim of the present paper is to consider the structure factor
of a dilute Bose-condensed system, in the frame of the self-consistent
mean-field approximation [21--26], and to analyze the properties of the
structure factor. This approximation, as is shown in Ref. [26], can be
employed for both weak as well as strong atomic interactions. Under the
Bose-condensed system, one can keep in mind atomic dilute Bose gases
[10--15], molecular systems formed by bound bosons [27--30] and fermions
[31--35], which can be achieved by invoking the Feshbach resonance [36,37],
or other effectively bosonic assemblies.

Throughout the paper the system of units is used, where the Planck
constant and Boltzmann constant are set to unity, $\hbar\equiv 1$,
$k_B\equiv 1$.

\section{Bose-condensed system}

Phase transitions are known to usually be accompanied by spontaneous
symmetry breaking [38]. The appearance of Bose-Einstein condensate is
associated with the breaking of global gauge symmetry. The most convenient
method of breaking the latter is by means of the Bogolubov shift [39,40].
Then, the field operator is represented in the shifted form
\be
\label{1}
\hat\psi(\br,t) \equiv \eta(\br,t) + \psi_1(\br,t) \; ,
\ee
in which $\eta(\br,t)$ is a nonoperator function, defining the condensate
wave function, and $\psi_1(\br,t)$ is the field operator of uncondensed
particles, satisfying the Bose commutation relations. The field variables
of condensed and uncondensed particles are orthogonal to each other,
\be
\label{2}
\int \eta^*(\br,t) \psi_1(\br,t) \; d\br = 0 \; .
\ee

The condensate function is normalized to the number of condensed particles
\be
\label{3}
N_0 = \int |\eta(\br,t)|^2 \; d\br \; ,
\ee
while the operator of uncondensed particles satisfies the normalization
\be
\label{4}
N_1 = <\hat N_1> \; ,
\ee
given by statistically averaging the operator
\be
\label{5}
\hat N_1 \equiv \int \psi_1^\dgr(\br,t) \psi_1(\br,t) \; d\br
\ee
for the number of uncondensed particles. The average number of particles
in the whole system is $N=N_0+N_1$. The statistical condition
\be
\label{6}
<\psi_1(\br,t)>\; = \; 0
\ee
guarantees the conservation of quantum numbers.

Gauge symmetry can also be broken by introducing infinitesimal sources
and defining the Bogolubov quasiaverages [39,40]. Both these methods of
gauge symmetry breaking, by means of the Bogolubov shift (1) and by means
of infinitesimal sources, have been shown to be equivalent and sufficient
for describing Bose-Einstein condensation [39--45].

The operators of observables, after the Bogolubov shift (1), are defined
on the Fock space ${\cal F}(\psi_1)$ generated by $\psi_1^\dgr$ (see
mathematical details in Refs. [23,46]). The energy operator is taken in the
standard form
\be
\label{7}
\hat H = \int \hat\psi^\dgr(\br) \left ( -\; \frac{\nabla^2}{2m} +
U \right ) \hat\psi(\br) \; d\br + \frac{1}{2} \; \int \hat\psi^\dgr(\br)
\hat\psi^\dgr(\br') \Phi(\br-\br') \hat\psi(\br') \hat\psi(\br)\; d\br\;
d\br' \; ,
\ee
in which, for breavity, the time dependence of the field operators is not
shown explicitly; $U=U(\br,t)$ is an external field, if any; and $\Phi(\br)$
is the interaction potential.

In what follows, we shall consider a uniform system, with no external
fields, $U=0$. The system is assumed to be equilibrium. An equilibrium
statistical system is to be characterized by a representative statistical
ensemble [47], which is the couple $\{ {\cal F}(\psi_1),\hat\rho\}$ of the
Fock space ${\cal F}(\psi_1)$ and the appropriate statistical operator
$\hat\rho$. The latter can be found from the minimization of the information
functional [46,47], incorporating the normalization conditions (3) and (4).
This yields [21--26] the statistical operator
\be
\label{8}
\hat\rho =
\frac{e^{-\bt H[\eta,\psi_1]}}{{\rm Tr} e^{-\bt H[\eta,\psi_1]}} \; ,
\ee
in which $\bt=1/T$ is inverse temperature and the grand Hamiltonian is
\be
\label{9}
H[\eta,\psi_1] = \hat H - \mu_0 N_0 - \mu_1 \hat N_1 \; ,
\ee
where $\mu_0$ and $\mu_1$ are the Lagrange multipliers guaranteeing the
validity of the normalization conditions (3) and (4).

For a uniform system, we may set
\be
\label{10}
\eta(\br,t) =\sqrt{\rho_0} \qquad \left ( \rho_0 \equiv
\frac{N_0}{V}\right ) \; ,
\ee
with $V$ being the system volume. From $\Phi(\br)$ and $\psi_1(\br)$, we
pass to their Fourier transforms $\Phi_{\bf k}$ and $a_{\bf k}$, respectively. The
lowest nonzero averages of the products of $a_{\bf k}$ are the normal average
\be
\label{11}
n_{\bf k} \equiv \; < a_{\bf k}^\dgr a_{\bf k} >
\ee
and the anomalous average
\be
\label{12}
\sgm_{\bf k} \equiv \; < a_{\bf k} a_{-{\bf k}} > \; .
\ee
The average $<a_{\bf k}>=0$ is zero, according to condition (6).

The higher orders of the products of $a_{\bf k}$ can be simplified in the
Hartree-Fock-Bogolubov approximation [22--26]. To this end, we define
\be
\label{13}
\om_{\bf k} \equiv \frac{k^2}{2m} + \rho \Phi_0 + \rho_0 \Phi_{\bf k} +
\frac{1}{V} \; \sum_{{\bf p}\neq 0} n_{\bf p} \Phi_{{\bf k}+{\bf p}} \; - \; \mu_1
\ee
and
\be
\label{14}
\Dlt_{\bf k} \equiv \rho_0 \Phi_{\bf k} + \frac{1}{V} \; \sum_{{\bf p}\neq 0}
\sgm_{\bf p} \Phi_{{\bf k}+{\bf p}} \; .
\ee
Then the grand Hamiltonian (9) acquires the form
\be
\label{15}
H_{HFB} = E_{HFB} + \sum_{{\bf k}\neq 0} \om_{\bf k} a_{\bf k}^\dgr a_{\bf k} + \;
\frac{1}{2} \; \sum_{{\bf k}\neq 0} \Dlt_{\bf k} \left ( a_{\bf k}^\dgr a_{-{\bf k}}^\dgr +
a_{-{\bf k}} a_{\bf k} \right ) \; ,
\ee
with the nonoperator part
\be
\label{16}
E_{HFB} =\left ( \frac{1}{2}\; \rho_0 \Phi_0  - \mu_0 \right ) N_0 -
\; \frac{1}{2}\; \rho_1^2 \Phi_0 V \; - \; \frac{1}{2V} \;
\sum_{{\bf k},{\bf p}\neq 0} \Phi_{{\bf k}+{\bf p}} ( n_{\bf k} n_{\bf p} +
\sgm_{\bf k}\sgm_{\bf p} ) \; ,
\ee
where $\rho_1\equiv N_1/V$.

The grand thermodynamic potential is
\be
\label{17}
\Om = - T \ln {\rm Tr} e^{-\bt H[\eta,\psi_1]} \; .
\ee
For thermodynamic stability, the number of condensed particles $N_0$ is
to be a minimizer of the grand potential (17), which implies
\be
\label{18}
\frac{\prt\Om}{\prt N_0} = \; <
\frac{\prt H[\eta,\psi_1]}{\prt N_0} > \; = \; 0 \; .
\ee
The latter defines the condensate chemical potential
\be
\label{19}
\mu_0 = \rho \Phi_0 + \frac{1}{V} \; \sum_{{\bf p}\neq 0} (n_{\bf p} +
\sgm_{\bf p}) \Phi_{\bf p} \; .
\ee

The grand Hamiltonian (15) can be diagonalized by means of the Bogolubov
canonical transformation
\be
\label{20}
a_{\bf k} = u_{\bf k} b_{\bf k} + v_{-{\bf k}}^* b_{-{\bf k}}^\dgr \; , \qquad
u_{\bf k}^2 = \frac{\om_{\bf k} +\ep_{\bf k}}{2\ep_{\bf k}} \; , \qquad
v_{\bf k}^2 = \frac{\om_{\bf k} -\ep_{\bf k}}{2\ep_{\bf k}} \; ,
\ee
with the Bogolubov spectrum
\be
\label{21}
\ep_{\bf k} =\sqrt{\om_{\bf k}^2 -\Dlt_{\bf k}^2} \; .
\ee
Then Hamiltonian (15) reduces to the diagonal form
\be
\label{22}
H_B = E_B + \sum_{{\bf k}\neq 0} \ep_{\bf k} b_{\bf k}^\dgr b_{\bf k} \; ,
\ee
in which
\be
\label{23}
E_B = E_{HFB} + \frac{1}{2}\; \sum_{{\bf k}\neq 0} (\ep_{\bf k} -\om_{\bf k}) \; .
\ee

The existence of Bose-Einstein condensate requires the condition
\be
\label{24}
\lim_{k\ra 0} \ep_{\bf k} = 0 \; , \qquad \ep_{\bf k} \geq 0 \; ,
\ee
from which the chemical potential of uncondensed particles follows as
\be
\label{25}
\mu_1 = \rho \Phi_0 + \frac{1}{V} \; \sum_{{\bf p}\neq 0} ( n_{\bf p} - \sgm_{\bf p})
\Phi_{\bf p} \; .
\ee
The Lagrange multipliers (19) and (25) are, evidently, different. They would
coincide only for the ideal gas, when $\Phi_{\bf k}\equiv 0$, hence $\mu_0=\mu_1=0$.
But for a nonideal system, $\mu_0\neq\mu_1$. Sometimes, one sets these
potentials equal by omitting the anomalous average $\sgm_{\bf p}$, which corresponds
to the Shohno model [48]. However, direct calculations [25,26,49] demonstrate
that the anomalous average is of the order and can even be larger than the
normal average. Therefore, omitting the larger quantity than what is retained
has nothing to do with a reasonable approximation. In addition, neglecting the
anomalous average makes the system unstable [21,50,51].

\section{Structure factor}

The structure factor is an important characteristic, which can be measured
in scattering experiments, such as neutron or photon scatterings [52]. Let
us consider, for concreteness, neutron scattering. A neutron, with the
initial energy $E_i$ and momentum $\bk_i$, scatters on an atom, acquiring
the final energy $E_f$ and momentum $\bk_f$. The momentum and energy,
transferred by a neutron to an atom, are $\bk\equiv\bk_i-\bk_f$ and
$\om\equiv E_i-E_f$. The double-differential scattering cross section of
neutrons on the system can be written [38] as
\be
\label{26}
\frac{1}{N} \; \frac{d^2\sgm}{d\Om_{\bf k} d\om} = b_s^2 \;
\frac{k_f}{k_i} \; S_{tot}(\bk,\om) \; ,
\ee
where $\Om_{\bf k}$ is a solid angle around $\bk$; $b_s$ is the scattering
length of the neutron on an atom; and $S_{tot}(\bk,\om)$ is the total
dynamic structure factor. The latter is defined by the expression
\be
\label{27}
S_{tot}(\bk,\om) = \frac{1}{N} \; \int R(\br,t,\br',0)\;
\exp\left \{ -i\bk (\br-\br') + i\om t\right \} \; d\br d\br' dt \; ,
\ee
in which
\be
\label{28}
R(\br,t,\br',t') \; \equiv \; <\hat n(\br,t) \hat n(\br',t') >
\ee
is the density-density correlation function, and the density operator is
\be
\label{29}
\hat n(\br,t) \equiv \hat\psi^\dgr(\br,t) \hat\psi(\br,t) \; .
\ee

The total dynamic structure factor (27) can be decomposed into two terms
\be
\label{30}
S_{tot}(\bk,\om) = S_e(\bk,\om) + S(\bk,\om) \; ;
\ee
the first term
\be
\label{31}
S_e(\bk,\om) \equiv \frac{1}{N} \; \int \rho(\br,t) \rho(\br',0) \;
\exp\left \{ -i \bk (\br-\br') + i\om t\right \} \; d\br d\br' dt \; ,
\ee
with $\rho(\br,t)\equiv<\hat n(\br,t)>$, is responsible for elastic
scattering and is called the disconnected dynamic structure factor; the
second term, called connected,
\be
\label{32}
S(\bk,\om) \equiv \frac{1}{N} \; \int \left [ R(\br,t,\br',0) -
\rho(\br,t) \rho(\br',0) \right ] \; \exp \left \{ -i \bk (\br-\br') + i\om
t\right \} \; d\br d\br' dt
\ee
describes inelastic scattering.

The elastic dynamic factor (31) is nontrivial for nonuniform systems [53,54],
but for a uniform equilibrium system, when $\rho(\br,t)=\rho$, it is very
simple, being
\be
\label{33}
S_e(\bk,\om) = (2\pi)^4 \rho \dlt(\bk) \; \dlt(\om) \; .
\ee
Therefore, for uniform systems, more attention is payed to the inelastic
dynamic factor (32). The latter enjoys the sum rules, one of which
\be
\label{34}
\int_{-\infty}^\infty \; S(\bk,\om) \; \frac{d\om}{2\pi} = S(\bk)
\ee
defines the static structure factor $S(\bk)$, another gives
\be
\label{35}
\int_{-\infty}^\infty \;\om S(\bk,\om) \; \frac{d\om}{2\pi} =
\frac{k^2}{2m} \; ,
\ee
and the third, connects the integral
\be
\label{36}
\int_{-\infty}^\infty \;\frac{1}{\om}\; S(\bk,\om) \; \frac{d\om}{2\pi} =
-\; \frac{{\rm Re}\chi(\bk,0)}{2\rho} \; ,
\ee
with the responce function $\chi(\bk,\om)$. At very large momentum
$\bk\ra\infty$, there are the following asymptotic equalities:
$$
\int_{-\infty}^\infty \; S(\bk,\om) \; \frac{d\om}{2\pi} \simeq 1
\qquad (k\ra \infty) \; ,
$$
$$
\int_{-\infty}^\infty \; \left ( \om -\; \frac{k^2}{2m} \right )
S(\bk,\om) \; \frac{d\om}{2\pi} \simeq 0 \; ,
$$
$$
\int_{-\infty}^\infty \; \left ( \om -\; \frac{k^2}{2m} \right )^2
S(\bk,\om) \; \frac{d\om}{2\pi} \simeq \frac{2k^2}{3m} \;
\frac{<\hat K>}{N} \; ,
$$
$$
\int_{-\infty}^\infty \; \left ( \om -\; \frac{k^2}{2m} \right )^3
S(\bk,\om) \; \frac{d\om}{2\pi} \simeq 0 \; ,
$$
here $<\hat K>$ is the average kinetic energy of the system.

For an equilibrium system, where
\be
\label{37}
\rho(\br,t) = \rho(\br) \; = \; <\hat n(\br) > \; ,
\ee
with $\hat n(\br)\equiv \hat n(\br,0)$, the static structure factor, defined
in Eq. (34), becomes
\be
\label{38}
S(\bk) = \frac{1}{N} \; \int \left [ <\hat n(\br) \hat n(\br')>  - \;
\rho(\br)\rho(\br')\right ]\; \exp\left \{ -ik (\br-\br')\right \} \;
d\br d\br' \; .
\ee
Let us introduce the Fourier transform of the density operator (29),
\be
\label{39}
\hat\rho_{\bf k} \equiv \int \hat n(\br)\; e^{-i\bk\cdot\br}\;
d\br \; , \qquad \hat n(\br) = \frac{1}{V} \; \sum_{\bf k} \hat \rho_{\bf k} \;
e^{i\bk\cdot\br} \; .
\ee
The following properties hold:
$$
\left [ \hat\rho_{\bf k} , \; \hat\rho_{\bf p} \right ] = 0 \; , \qquad
\left [ \hat n(\br) , \; \hat n(\br') \right ] = 0 \; , \qquad
\hat\rho_{\bf k}^+ = \hat\rho_{-{\bf k}} \; , \qquad \hat n^+(\br) =  n(\br) \; .
$$
The zero component of $\rho_{\bf k}$ yields the total number of particles,
\be
\label{40}
\hat\rho_0 = \int \hat n(\br)\; d\br = \hat N \; .
\ee
The static structure factor (38) can be expressed as
\be
\label{41}
S(\bk) = \frac{\Dlt^2(\hat\rho_{\bf k})}{N}
\ee
through the dispersion
\be
\label{42}
\Dlt^2(\hat\rho_{\bf k}) \; \equiv \; <\hat\rho_{\bf k}^+ \hat\rho_{\bf k}> -
<\hat\rho_{\bf k}^+>< \hat\rho_{\bf k} > \; .
\ee

The central value of the structure factor (41),
\be
\label{43}
S(0) = \frac{\Dlt^2(\hat N)}{N} = \rho T \kappa_T = \frac{T}{ms^2}
\ee
connects it with the particle dispersion $\Dlt^2(\hat N)=\Dlt^2(\hat\rho_0)$,
the isothermal compressibility
\be
\label{44}
\kappa_T \equiv -\; \frac{1}{V} \left ( \frac{\prt V}{\prt P}
\right )_{TN} =  \frac{1}{\rho} \left ( \frac{\prt\rho}{\prt P}
\right )_{TN} \; ,
\ee
where $P$ is pressure, and with the hydrodynamics sound velocity $s$,
\be
\label{45}
s^2 \equiv \frac{1}{m} \left ( \frac{\prt P}{\prt\rho} \right )_{T} =
\frac{1}{m\rho\kappa_T} \; .
\ee
Thus, the isothermal compressibility and sound velocity are expressed
through the structure factor as
\be
\label{46}
\kappa_T = \frac{S(0)}{\rho T} \; , \qquad s^2 = \frac{T}{mS(0)} \; .
\ee
Relations (43) and (46) are exact, being valid for any equilibrium, not
necessary uniform, system.

Passing to a uniform system, for the Bogolubov shift (1), we have
\be
\label{47}
\hat\psi(\br) = \sqrt{\rho_0} \; + \; \psi_1(\br) \; .
\ee
The density operator (29) becomes
\be
\label{48}
\hat n(\br) = \rho_0 + \hat n_1(\br) +
\sqrt{\rho_0}\left [ \psi_1^\dgr(\br) + \psi_1(\br)\right ] \; ,
\ee
where the density operator of uncondensed particles is
\be
\label{49}
\hat n_1(\br) \equiv \psi_1^\dgr(\br) \psi_1(\br) \; .
\ee
The Fourier transform of the total density operator (48) takes the form
\be
\label{50}
\hat\rho_{\bf k} = \dlt_{{\bf k}0} N_0 + \sqrt{N} \left ( a_{-{\bf k}}^\dgr +
a_{\bf k} \right ) + \sum_{{\bf p}\neq 0} a_{{\bf k}+{\bf p}}^\dgr a_{\bf p} \; .
\ee
Because of the properties
$$
<a_{\bf k}^\dgr a_{\bf p}> \; = \; \dlt_{{\bf kp}} n_{\bf k} \; , \qquad
<a_{\bf k}> \; = \; 0 \; ,
$$
we have
\be
\label{51}
<\hat\rho_{\bf k} > \; = \; \dlt_{{\bf k}0} N \; .
\ee
For dispersion (42), we find
\be
\label{52}
\Dlt^2(\hat\rho_{\bf k}) = N + 2N_0 ( n_{\bf k} + \sgm_{\bf k}) +
\sum_{{\bf p},{\bf q}\neq 0} <a_{\bf p}^\dgr
a_{{\bf k}+{\bf q}}^\dgr a_{{\bf k}+{\bf p}} a_{\bf q} > - \;
\dlt_{{\bf k}0} N_1^2 \; .
\ee

When the interaction potential $\Phi(\br)$ is singular at $r\ra 0$, then
pair atomic correlations are very important [52,55]. But here, we have
assumed that the interaction potential is soft, so that it possesses the
Fourier transform $\Phi_{\bf k}$. For the soft potential, we can apply the
Hartree-Fock-Bogolubov (HFB) approximation. Employing the latter, one has
to be cautious, since its blind application can lead to unphysical results.
Thus, if we would formally decouple the four-operator term in dispersion
(52), by using the HFB approximation, we would come to
\be
\label{53}
\Dlt^2(\hat\rho_{\bf k}) \longrightarrow N + 2N_0 (n_{\bf k} +\sgm_{\bf k}) +
\sum_{{\bf p}\neq 0} (n_{{\bf k}+{\bf p}} n_{\bf p} + \sgm_{{\bf k}+{\bf p}}
\sgm_{\bf p} ) \; .
\ee
However, one should keep in mind that the use of the HFB approximation
results in the Hamiltonian (22), which is quadratic with respect to the
operators of uncondensed particles. Therefore, all expressions of the order
higher than two with respect to these operators are not well defined in this
approximation. One could consider such high-order terms only if their
behavior were reasonable. But sometimes these higher-order terms result in
unphysical divergencies, which simply means that they should be omitted,
being outside of the region of applicability of the used approximation. In
the second-order approximation, only the second-order terms are well defined
[14,21,23,50,51]. Consequently, dispersion (52), in the HFB approximation,
has to be written as
\be
\label{54}
\Dlt^2(\hat\rho_{\bf k}) = N + 2N (n_{\bf k} +\sgm_{\bf k}) \; .
\ee
Recall that the normal and anomalous averages in the HFB approximation
[23--26] are
\be
\label{55}
n_{\bf k} = \frac{\om_{\bf k}}{2\ep_{\bf k}} \; {\rm coth}\left ( \frac{\ep_{\bf k}}{2T}\right )
- \; \frac{1}{2} \; , \qquad \sgm_{\bf k} = -\; \frac{\Dlt_{\bf k}}{2\ep_{\bf k}} \;
 {\rm coth}\left ( \frac{\ep_{\bf k}}{2T}\right ) \; ,
\ee
where $\om_{\bf k}$, $\Dlt_{\bf k}$, and $\ep_{\bf k}$ are given by Eqs. (13), (14), and (21),
respectively. The Bogolubov spectrum (21), according to condition (24), is
gapless, having the phonon form $\ep_{\bf k}\simeq ck$ at $k\ra 0$, with the sound
velocity $c$ defined by the equation
\be
\label{56}
mc^2 = \rho_0 \Phi_0 + \frac{1}{V}\; \sum_{{\bf p}\neq 0} \sgm_{\bf p} \Phi_{\bf p} \; .
\ee

Dispersion (54) is finite. For instance, in the case of $k\ra 0$, when
$\Dlt^2(\hat\rho_0)=\Dlt^2(\hat N)$, we have
$$
\Dlt^2(\hat N) = N \left [ 1 + 2 \lim_{k\ra 0} ( n_{\bf k} +\sgm_{\bf k})
\right ] \; .
$$
Using Eqs. (55), we get
$$
\lim_{k\ra 0} (n_{\bf k} +\sgm_{\bf k}) = \frac{1}{2}\left ( \frac{T}{mc^2} \; - \;
1\right ) \; .
$$
Therefore,
\be
\label{57}
\Dlt^2(\hat N) = \frac{NT}{mc^2} \; .
\ee
Consequently,
\be
\label{58}
S(0) = \frac{T}{mc^2} \; , \qquad \kappa_T = \frac{1}{\rho mc^2} \; ,
\qquad s = c \; .
\ee

The structure factor (41), according to Eq. (54), takes the form
\be
\label{59}
S(\bk) = 1 + 2(n_{\bf k} +\sgm_{\bf k}) \; .
\ee
Substituting here expressions (55), we find
\be
\label{60}
S(\bk) = \frac{k^2}{2m\ep_{\bf k}} \; {\rm coth}\left (
\frac{\ep_{\bf k}}{2T}\right ) \; .
\ee
Note that the Feynman relation
$$
\ep_{\bf k} = \frac{k^2}{2m S(\bk)} \qquad (T=0)
$$
is valid only at zero temperature. The asymptotic properties of the
structure factor (60) are:
$$
S(\bk) \simeq \frac{T}{mc^2} \qquad (k\ra 0 ) \; ,
$$
$$
S(\bk) \simeq 1 \qquad (k\ra \infty) \; .
$$

If instead of expression (54), one would use Eq. (53), then the structure
factor $S(\bk)$ would be divergent, which has no physical sense. Hence, Eq.
(54) provides the correct mean-field approximation. The correctness of Eq.
(54) will be analyzed in more detail in the following section.

\section{Approximation analysis}

The form of the structure factor (60) is due to that of dispersion (54).
Here, we shall analyze the correctness of the approximation in which Eq.
(53) has been replaced by Eq. (54).

We shall also stress the necessity of correctly taking into account the
anomalous averages and the importance of gauge symmetry breaking.

\subsection{Preservation of approximation order}

The replacement of Eq. (53) by Eq. (54) is based on the requirement of
preserving the approximation order with respect to the operators of
uncondensed particles. Suppose that we would go outside of the region of
applicability of the considered second-order approximation and would retain
the last term in Eq. (53). Then $\Dlt^2(\hat\rho_{\bf k})$ would be proportional
to $N^{4/3}$, hence $S(\bk)$ proportional to $N^{1/3}$, that is, the
structural factor $S(\bk)$ would be divergent, hence, unphysical. To show
this, consider $\Dlt^2(\hat\rho_0)=\Dlt^2(\hat N)$. Then the last terms in
Eq.
(53) give
$$
V \int n_{\bf k}^2 \; \frac{d\bk}{(2\pi)^3} \sim \frac{(mT)^2}{2\pi^2}\;
V^{4/3} \; , \qquad
V \int \sgm_{\bf k}^2 \; \frac{d\bk}{(2\pi)^3} \sim \frac{(mT)^2}{2\pi^2}\;
V^{4/3} \; .
$$
Therefore,
$$
\Dlt^2(\hat N) \sim \frac{(mT)^2}{\pi}\; V^{4/3} \; .
$$
In this way, when $T>0$, then in thermodynamic limit
$$
S(0) \; \propto \; N^{1/3} \; \ra \; \infty \; , \qquad
s  \; \propto \; N^{-1/6} \; \ra \; 0 \; ,
$$
$$
\kappa_T  \; \propto \; N^{1/3} \; \ra \;  \infty  \qquad
(N \ra \infty) \; .
$$
This means an absolute instability  of the system at any finite temperature.

At zero temperature, Eqs. (55) reduce to
$$
n_{\bf k} = \frac{1+2q^2}{4q\sqrt{1+q^2}} \; - \; \frac{1}{2} \; ,
\qquad \sgm_{\bf k} = -\; \frac{1}{4q\sqrt{1+q^2}}  \qquad (q \equiv
\frac{k}{2mc} \; , \; T=0 ) \; .
$$
From here,
$$
\lim_{k\ra 0} (n_{\bf k} + \sgm_{\bf k}) = -\; \frac{1}{2} \qquad (T=0) \; .
$$
The last terms in Eq. (53), at $T=0$, are finite, since
$$
\int  n_{\bf k}^2 \; \frac{d\bk}{(2\pi)^3} < \infty \; , \qquad
\int  \sgm_{\bf k}^2 \; \frac{d\bk}{(2\pi)^3} < \infty \; .
$$
The dispersion $\Dlt^2(\hat N)\propto N$ is thermodynamically normal. The
value $S(0)\propto const >0$ is finite. However, as follows from relations
(46),
$$
\kappa_T \; \propto \; \frac{1}{T} \; \ra \; \infty \; , \qquad
s \; \propto \; T \; \ra \; 0 \qquad (T\ra 0) \; ,
$$
which tells that the system is again unstable. Thus, at any temperature
$T\geq 0$, in thermodynamic limit, one has $s=0$ and $\kappa_T=\infty$.
Hence, such a system would always be unstable. But, of course, the appearance
of this instability is not physical and is caused only by going outside of
the region of applicability of the used approximation. Preserving the
approximation order requires to reject the last terms in Eq. (53). Then no
unphysical divergencies appear.

\subsection{Importance of anomalous averages}

One of the popular tricks when considering Bose-condensed systems is the
omission of the anomalous average $\sgm_{\bf k}$, which corresponds to the Shohno
model [48]. One often ascribes this unjustified trick to Popov. But, as easy
to check from the original works by Popov [56,57], which are cited in this
respect, he has never suggested this trick. Here we show that omitting the
anomalous average is principally wrong, since then the system becomes unstable
and its structural factor infinite, which is, certainly, of no sense.

Assume that the anomalous average $\sgm_{\bf k}$ in Eq. (54) is set to zero. Then
we have
$$
\Dlt^2(\hat N) \ra N \left ( 1 + 2 \lim_{k\ra 0} n_{\bf k} \right ) \; .
$$
At nonzero temperature,
$$
n_{\bf k} \simeq \frac{mT}{k^2} \qquad (T>0, \; k\ra 0) \; .
$$
Since $k\sim 1/V^{1/3}$, we get
$$
\Dlt^2(\hat N) \sim \frac{2mT}{\rho^{2/3}} \; N^{5/2} \qquad
(T>0) \; .
$$
Then, at finite temperature,
$$
S(0) \sim \frac{2mT}{\rho^{2/3}} \; N^{2/3} \; , \qquad
\kappa_T \sim \frac{2m}{\rho^{5/3}} \; N^{2/3} \; .
$$
These quantities diverge in thermodynamic limit, which implies instability.

At zero temperature,
$$
n_{\bf k} \simeq \frac{mc}{2k} \qquad (T=0, \; k\ra 0) \; .
$$
From here
$$
\Dlt^2(\hat N) \sim \frac{mc}{\rho^{1/3}} \; N^{4/3} \qquad
(T=0) \; .
$$
Hence, as $T\ra 0$, we have
$$
S(0) \sim \frac{mc}{\rho^{1/3}} \; N^{1/3} \; , \qquad \kappa_T \sim
\frac{mc}{\rho^{4/3}T} \; N^{1/3} \; .
$$
The structural factor diverges in thermodynamic limit, and the isothermal
compressibility $\kappa_T$ becomes divergent even at finite $N$, when $T\ra
0$.

Thus, at all temperatures $T\geq 0$, we come to the senseless unphysical
behavior, when
$$
S(\bk) \ra \infty \; , \qquad s \ra 0\; , \qquad \kappa_T \ra \infty \; .
$$
This proves that the anomalous average plays a principally important role
and cannot be omitted.

\subsection{Gauge symmetry breaking}

Gauge symmetry breaking is known to be {\it sufficient} for describing
Bose-Einstein condensation. This follows from the general theorems,
independent from approximations, proved by Bogolubov [39,40], Ginibre
[41], and Lieb et al. [42--45]. At the same time, in the Penrose-Onsager
[58] scheme of Bose condensation no explicit breaking of symmetry is
required, but the appearance of condensate is manifested in the properties
of the density-matrices eigenvalues or their order indices [59,60]. Here we
give arguments that, in the frame of a mean-field approximation, gauge
symmetry breaking is {\it necessary}.

Suppose, we consider a mean-field approximation without breaking the gauge
symmetry. Then the number-of-particle dispersion has the standard form
$$
\Dlt^2(\hat N) = \sum_{\bf k} n_{\bf k}(1+n_{\bf k}) = \Dlt^2(\hat N_0) +
\Dlt^2(\hat N_1) \; ,
$$
which can be separated into two terms,
$$
\Dlt^2(\hat N_0) = N_0 ( 1+ N_0) \; , \qquad
\Dlt^2(\hat N_1) =  \sum_{{\bf k}\neq 0} n_{\bf k}(1+n_{\bf k}) \; ,
$$
corresponding to condensed and uncondensed particles, respectively. The
occurrence of Bose-Einstein condensation implies that the number of
condensed particles becomes proportional to the total number of particles,
$N_0\propto N$. As a result,
$$
\Dlt^2(\hat N_0) \; \propto \; N^2 \qquad (N_0  \propto N) \; .
$$
Consequently, $\Dlt^2(\hat N)\propto N^2$. Therefore, the isothermal
compressibility and the structure factor diverge in the thermodynamic limit
as
$$
S(\bk)  \; \propto \; N \; \ra \; \infty \; , \qquad
\kappa_T \; \propto \; N \; \ra \; \infty \; ,
$$
which means instability.

In this way, the use of the mean-field approximation requires that gauge
symmetry be broken. It may be that the exact consideration or some refined
approximations (see, e.g., Ref. [17]) allow for a description without gauge
symmetry breaking. However, for treating Bose-Einstein condensation in a
mean-field picture, gauge symmetry breaking is the necessary and sufficient
condition. This also concerns any perturbation theory starting with a
mean-field approximation.

\section{Conclusion}

A Bose system with Bose-Einstein condensate is considered. The properties
of the structure factor are investigated in the frame of the self-consistent
mean-field approximation, which is both conserving and gapless. The completed
analysis demonstrates the importance of preserving the approximation order.
Going outside of the region of applicability of the used approximation can
lead to the appearance of unphysical divergences. In the system with broken
gauge symmetry, it is crucially important to accurately take account of
anomalous averages. Omitting the latter is unjustified and results in the
occurrence of fictitious instability. It is also shown that gauge symmetry
breaking is necessary in any theory based on a mean-field picture.

\end{document}